\documentclass[12pt]{article}
\usepackage[latin1]{inputenc}
\usepackage{amsmath}
\usepackage{amsfonts}
\usepackage{amssymb}
\usepackage{graphicx}
\usepackage{footnote}
\usepackage{float}

\usepackage{subfigure}

\DeclareGraphicsExtensions{.bmp,.png,.pdf,.jpg,.eps}

\usepackage[colorlinks]{hyperref}
\hypersetup{
    colorlinks=true,
    linkcolor=blue,
    filecolor=magenta,  
    urlcolor=[rgb]{0.00,0.00,0.50},    
    citecolor=red,
    }

\usepackage{url}

\advance \textheight by \topskip
\usepackage{amsmath}
\usepackage[english]{babel}
\makeatletter
 \@addtoreset{equation}{section}
\makeatother

\usepackage{amsmath}
\numberwithin{equation}{section}

\advance \textheight by \topskip
\usepackage{amsmath}
\usepackage[english]{babel}

\def\be{\begin{equation}}
\def\ee{\end{equation}}

\def\A{\mathbb A}
\def\B{\mathbb B}
\def\Z{\mathbb Z}
\def\R{\mathbb R}
\def\W{\mathbb W}

\usepackage{todonotes}

\begin{document}
%%%%%%%%%%%%%%%%%%%%%%%%%%%%%

\title{
{\bf 
Nonlinear supersymmetry as a\\ hidden symmetry
 }}

\author{{\bf Mikhail S. Plyushchay} 
 \\
[8pt]
{\small \textit{ Departamento de F\'{\i}sica,
Universidad de Santiago de Chile, Casilla 307, Santiago,
Chile  }}\\
[4pt]
 \sl{\small{E-mail:  
\textcolor{blue}{mikhail.plyushchay@usach.cl}
}}
}
\date{}
\maketitle

\begin{abstract}
Nonlinear supersymmetry  
is characterized by supercharges to be higher order 
in bosonic momenta of a system, and thus has a nature of a  hidden symmetry.
We review some aspects of  nonlinear supersymmetry
and related  to it
exotic supersymmetry and nonlinear superconformal symmetry.
Examples of reflectionless, finite-gap and  
perfectly invisible $\mathcal{PT}$-symmetric 
zero-gap systems,  as well as 
rational deformations of the  quantum harmonic oscillator and conformal mechanics, 
are considered, in which  such  symmetries are realized.
\end{abstract}

\vskip.5cm\noindent

%%%%%%%%%%%%%%%%%%%%%%%%%%%%%%%%%%%%%%%%%%%%%%%%
\section{Introduction}\label{SecIntro}
%%%%%%%%%%%%%%%%%%%%%%%%%%%%%%%%%%%%%%%%%%%%%%%%
Hidden symmetries  are associated with nonlinear in momenta  integrals of motion.  
They mix the coordinate and momenta variables in the phase space of a system,
and generate  a nonlinear, $W$-type algebras \cite{Cariglia}.
The best known examples  of hidden symmetries are
provided by the Laplace-Runge-Lenz vector integral in the Kepler-Coulomb 
problem, and the Fradkin-Jauch-Hill tensor in isotropic harmonic oscillator
systems.
Hidden symmetries also appear 
in anisotropic oscillator with conmensurable frequencies, 
where they
underlie  the closed nature of  classical trajectories
and 
specific degeneration of the quantum energy levels.
Hidden symmetry is responsible for complete integrability of geodesic
motion of a test particle in the background of the vacuum solution to the Einstein's
equation represented by the Kerr metric of the rotating black hole
and its generalizations in the form of 
the Kerr-NUT-(A)dS solutions of the Einstein-Maxwell
 equations \cite{Frolov}. 
 Another class of hidden symmetries underlies a  complete integrability of the field systems 
described by nonlinear wave equations such as the Korteweg-de Vries (KdV) equation. 
Those symmetries are responsible for peculiar properties of the soliton and 
finite-gap solutions of the KdV  system, whose equation
of  motion can be regarded  as a  geodesic flow 
on the  Virasoro-Bott group \cite{Khesin1,Khesin2}. 
\vskip0.1cm

Nonlinear supersymmetry   \cite{Dubov1}--\cite{BMP}
is characterized by supercharges to be higher order 
in even (bosonic) momenta of a system, and thus has a nature of hidden symmetry.
Here, we review some aspects of  nonlinear supersymmetry,
and related  to it
exotic supersymmetry and nonlinear superconformal symmetry. 
\vskip0.1cm

Nonlinear supersymmetry appears, particularly,  in purely
parabosonic harmonic oscillator systems 
generated by the deformed Heisenberg algebra with reflection 
\cite{PlyPara} as well as in a generalized Landau problem \cite{KliPly1}.
The peculiarity of supersymmetric parabosonic systems  shows up in
the nonlocal nature of  supercharges 
to be of infinite order in the momentum operator as well as in the ladder 
operators but  anti-commuting
for a polynomial in Hamiltonian  being quadratic in creation-annihilation 
operators. Similar peculiarities characterize 
hidden supersymmetry and hidden superconformal symmetry
appearing in some usual  quantum bosonic systems with a local Hamiltonian
operator \cite{Ply1,Ply2,Ply3,LeiPly,AnabPly,COP,CorPly,CorNiePly,CJNP,CJPtri,CorJakPly,JNPHid,CarPly1,InzPlyHid,InzuPlyHidconf}.
Exotic supersymmetry  emerges  in superextensions 
of the quantum systems described by soliton and finite-gap potentials,
in  which the key role is played by the Lax-Novikov integrals of motion 
\cite{CJNP,CJPtri,CorJakPly,CorrDunPl,ArMatPly1}.
A  structure   similar to that of the exotic supersymmetry  of reflectionless and finite-gap 
quantum systems can also be identified  in  the ``SUSY in the sky" type 
supersymmetry \cite{SUSYSky,Tanimoto,SUSYMono,PlyMono}
based on the presence of the Killing-Yano tensors in the mentioned above class of the 
black hole solutions to the  Einstein-Maxwell equations.
Nonlinear superconformal symmetry appears in rational deformations
of the quantum harmonic oscillator and conformal mechanics  
systems \cite{CarPly1,InzuPlyHidconf}. Both exotic supersymmetry and nonlinear superconformal 
symmetry characterize  the interesting class of 
the perfectly  invisible zero-gap $\mathcal{PT}$-symmetric  systems,
which includes the $\mathcal{PT}$-regularized two particle Calogero systems  and their rational extensions
with potentials satisfying the equations of the KdV hierarchy and exhibiting  a behaviour
of extreme (rogue) waves  \cite{MatPly,JM2}. 

\section{Nonlinear supersymmetry and  quantum anomaly}

Classical analog of the Witten's supersymmetric quantum mechanics
\cite{Witten1,Witten2,Cooper,Junker}
is described by the Hamiltonian 
\be\label{H1}
\mathcal{H}=
p^2+W^2
+W'N\,,
\ee
where $N=
\theta^+\theta^--\theta^-\theta^+
$, $W=W(x)$ is a superpotential, 
$x$ and $p$ are even canonical variables, $\{x,p\}=1$,
and $\theta^+$, $\theta^-=(\theta^+)^*$ are 
Grassmann variables with the only nonzero Poisson bracket
$\{\theta^+,\theta^-\}=-i$.
System (\ref{H1}) is characterized by the even,
$N$, and odd, $Q_+=(W+ip)\theta^+$
and $Q_-=(Q_+)^*$, integrals of motion
satisfying the algebra of $\mathcal{N}=2$ Poincar\'e supersymmetry
\be
\{Q_+,Q_-\}=-i\mathcal{H}\,,\quad
 \{\mathcal{H},Q_\pm\}=0\,,\quad
\{N,\mathcal{H}\}=0\,,\quad 
\{N,Q_\pm\}=\pm 2 iQ_\pm\,.
\ee
For any choice of the superpotential,
canonical quantization of this classical system 
gives rise to the supersymmetric quantum system
in which  
quantum supercharges and Hamiltonian
satisfy the $\mathcal{N}=2$ superalgebra 
given by a direct quantum analog 
of the corresponding Poisson bracket relations,
with the quantum analog of the integral $N$   
playing simultaneously the role of the $\Z_2$-grading operator
$\Gamma=\sigma_3$
of the Lie superalgebra.
\vskip0.1cm

A simple change of the last term in (\ref{H1}) for
$n\mathcal{W}'N$ with $n$ taking any integer value  yields  a  
system characterized by a nonlinear supersymmetry 
of order $n$ generated by the supercharges
 $S_+=(\mathcal{W}+ip)^n\theta^+$ and 
$S_-=(S_+)^*$ being the integrals of order
$n$ in the momentum $p$. Their
Poisson bracket
$\{S_+,S_-\}=-i(\mathcal{H})^n$ has order $n$ 
in the Hamiltonian \cite{PlyPara,KliPly,SchwAn}
\be\label{Hn}
\mathcal{H}=p^2+\mathcal{W}^2+n\mathcal{W}'N\,.
\ee
System (\ref{Hn}) can be regarded as a kind of the classical supersymmetric 
analog of the planar anisotropic oscillator with commensurable  
frequencies \cite{BDKL,BoerHarTji}.
Unlike a linear case (\ref{H1}) with $n=1$, canonical quantization of the system (\ref{Hn})
with $n=2,3,\ldots$
faces, however,   the problem of 
quantum anomaly: for arbitrary form of the superpotential,
quantum analogs of the classical odd integrals $S_\pm$ cease to commute
with the quantum analog of the Hamiltonian (\ref{Hn}).
In \cite{KliPly}, it was found a certain class of superpotentials $\mathcal{W}(x)$ for which
the supercharge $S_+$  has a  polynomial  structure in 
$z=\mathcal{W}+ip$ instead of monomial one 
so that the  corresponding systems  admit an anomaly-free quantization
giving rise to quasi-exactly solvable  systems 
\cite{Quasi1,Quasi2,Quasi3}.

If instead of the ``holomorphic'' dependence of the supercharge $S_+$
on the complex variable $z$ we consider  
the supercharges with polynomial dependence on the momentum variable $p$,
the case of quadratic supersymmetry turns out to be a special one.
The Hamiltonian and supercharges then can be presented in the most general form
\be\label{H2p}
\mathcal{H}=
zz^*-\frac{C}{\mathcal{W}^2}+4\mathcal{W}'N +a
\,,
\ee 
\be\label{Q2p}
S_+=
\left(
z^2+\frac{C}{\mathcal{W}^2}\right)
\theta^+\,,\qquad
S_-=(S_+)^*\,.
\ee
Here $a$ and $C$ are real constants,
and we have
\be\label{QQH2p}
\{S_+,S_-\}=-i\left((\mathcal{H}-a)^2+C\right)\,.
\ee 
Supersymmetry  of the system  (\ref{H2p}), (\ref{Q2p}), 
(\ref{QQH2p}) with an arbitrary superpotential 
can be preserved at the quantum level if to correct the direct 
quantum analog of the Hamiltonian and supercharges by  adding to them the
 term quadratic in Plank constant \cite{KliPly,SchwAn}:
\be\label{HDel}
\hat{\mathcal{H}}-a=-\hbar^2\frac{d^2}{dx^2}+\mathcal{W}^2-2\hbar\sigma_3\mathcal{W}'-
\frac{C}{\mathcal{W}^2}+\Delta(\mathcal{W})\,,
\ee
\be\label{QDel}
\hat{S}_+=\hat{s}_+\sigma_+\,,\qquad
\hat{s}_+=\left(\hbar\frac{d}{dx}+\mathcal{W}\right)^2+\frac{C}{\mathcal{W}^2}-
\Delta(\mathcal{W})\,,
\ee
\be
\Delta(\mathcal{W})=\frac{1}{2}\hbar^2 \left(
\frac{\mathcal{W}''}{\mathcal{W}}-\frac{1}{2}\left(\frac{\mathcal{W}'}{\mathcal{W}}\right)^2\right)=
\hbar^2\frac{1}{\sqrt{\mathcal{W}}}\left(\sqrt{\mathcal{W}}\right)''\,,
\ee
where $\sigma_+=\frac{1}{2}(\sigma_1+i\sigma_2)$.  
The quantum term $\Delta(\mathcal{W})$ can be presented
as a Schwarzian, $\Delta=-\frac{1}{2}\hbar^2S(\omega(x))$,
$S(\omega(x))=(\omega''/\omega')'-\frac{1}{2}(\omega''/\omega')^2$, 
of the  function $\omega(x)=\int^x dy/\mathcal{W}(y)$.
The quadratic in $\hbar$ terms in the quantum Hamiltonian  
(\ref{HDel}) can be unified and presented in a form  similar to that of the kinetic term 
of  the quantum particle in a curved space:
$-\hbar^2\frac{d^2}{dx^2}+\Delta(\mathcal{W})=\hat{\mathcal{P}}^\dagger \hat{\mathcal{P}}$,
where $\hat{\mathcal{P}}=\hbar \zeta^{-1}\frac{d}{dx}\zeta$, $\zeta=1/\sqrt{\mathcal{W}}$.
Analogously, the first and third terms in $\hat{s}_+$ in (\ref{QDel}) 
can be collected  and presented in the form  $\hat{z}^2-\Delta(W)=(\zeta \hat{z}\zeta^{-1})(\zeta^{-1}\hat{z}\zeta)$,
where $\hat{z}=\hbar \frac{d}{dx}+\mathcal{W}$  \cite{SchwAn}.

\section{Exotic nonlinear $\mathcal{N}=4$ supersymmetry}

The anomaly-free prescription for quantization of the classical systems (\ref{Hn})
with supersymmetry of order higher than two  in general case is unknown, but 
there exist infinite families of the quantum systems described by  
supersymmetries of arbitrary order.
They can be generated easily by applying 
the higher order Darboux-Crum (DC) transformations 
\cite{MatSal,Krein,Adler}
to a given, for instance,
 exactly solvable 
quantum  system
instead of starting from a classical supersymmetric system of the form (\ref{Hn})
followed  by a search for  the anomaly-free quantization scheme.

\vskip0.1cm
In general case the DC transformation 
of a given system 
described by the Hamiltonian operator $\hat{H}_-=-\frac{d^2}{dx^2}+V_-(x)$ 
is generated 
by selection of the set of physical or non-physical eigenstates $(\psi_1,\psi_2,\ldots,\psi_n)$
of $\hat{H}_-$ as the seed states. 
Here and below we put $\hbar=1$.
If they are chosen  in such a way that
their Wronskian $\W(\psi_1,\ldots,\psi_n)$ takes non-zero values in the  
region where  $V_-(x)$ is defined,  then the new potential 
\be\label{V-V+DC}
V_+=V_--2(\ln \W(\psi_1,\ldots,\psi_n))''
\ee
will be regular in the same region as $V_-$.
Physical and non-physical eigenstates of the new Hamiltonian
operator $\hat{H}_+=-\frac{d^2}{dx^2}+V_+$ 
are obtained  from those of the original system $\hat{H}_-$
by the transformation
\be\label{Anpsi}
\psi_{+,\lambda}=\frac{ \W(\psi_1,\ldots,\psi_n,\psi_\lambda)}{\W(\psi_1,\ldots,\psi_n)}=
\A_n\psi_\lambda\,,
\ee
where $\psi_\lambda$ is an eigenstate of $\hat{H}_-$ different from 
eigenstates in the set of the seed states with eigenvalue $E_\lambda\neq E_j$, $j=1,\ldots, n$.
The state $\psi_{+,\lambda}$ is of the same eigenvalue of $\hat{H}_+$
as $\psi_\lambda$ of $\hat{H}_-$, $ \hat{H}_-\psi_\lambda=\lambda\psi_\lambda\,\Rightarrow\,
\hat{H}_+\psi_{+,\lambda}=\lambda\psi_{+,\lambda}$,
and vise versa,  from $\hat{H}_+\psi_{+,\lambda}=\lambda\psi_{+,\lambda}$
it follows that $ \hat{H}_-\psi_\lambda=\lambda\psi_\lambda$.
Operator $\A_n$ in (\ref{Anpsi}) is a differential operator of order $n$,
\be\label{Andef}
\A_n=A_n\ldots A_1\,,\quad
 A_j=(A_{j-1}\psi_j)\frac{d}{dx} (A_{j-1}\psi_j)^{-1}\,,\quad
j=1,\ldots,n\,,\quad A_0=1\,,
\ee
which is constructed recursively from the selected seed states.
Operators $\A_n$ and $\A_n^\dagger$
intertwine Hamiltonian operators $\hat{H}_-$ and $\hat{H}_+$,
 \be\label{Ainter}
 \A_n\hat{H}_-=\hat{H}_+\A_n\,, \qquad 
 \A_n^\dagger\hat{H}_+=\hat{H}_-\A_n^\dagger\,,
 \ee
 and satisfy relations 
 \be\label{Aprod}
 \A_n^\dagger\A_n=\prod_{j=1}^{n}(\hat{H}_--E_j)\,,\qquad
 \A_n\A_n^\dagger=\prod_{j=1}^{n}(\hat{H}_+-E_j)\,,
 \ee
 where $E_j$ is  eigenvalue of the seed eigenstate $\psi_j$.
 Relations (\ref{Ainter}) and (\ref{Aprod}) underlie 
 nonlinear supersymmetry 
 of the extended system $\hat{\mathcal{H}}=\text{diag}\,(\hat{H}_+,\hat{H}_-)$,
 the supercharges of which are constructed from the operators
 $ \A_n$ and $ \A_n^\dagger$. 
 
 {}Using Eq. (\ref{Anpsi}), one can prove the relation \cite{CIPConf}
\be\label{Wtiltil}
\W(\psi_*,\widetilde{\psi_*},\psi_1,\ldots,\psi_n)=
\W(\psi_1,\ldots,\psi_n)\,.
\ee
Here and in what follows equality between 
Wronskians is implied up to inessential multiplicative constant;
$\psi_*$ is some eigenstate of $\hat{H}_-$ with eigenvalue $E_*$ different from 
$E_j$, $j=1,\ldots, n$, and $\widetilde{\psi_*}=\psi_*\int^x dy/(\psi_*(y))^2$
is a linear independent eigenstate with the same eigenvalue $E_*$ so that
$\W(\psi_*,\widetilde{\psi_*})=1$.
\vskip0.1cm

Among supersymmetric quantum systems generated by DC transformations, 
 there exists special class of infinite subfamilies  
in which the corresponding superextended systems are characterized simultaneously 
by supersymmetries of two different orders, one of which is of even order $n=2l$,
while another has some odd order $n=2k+1$ 
\cite{CJNP,CJPtri,CorJakPly,PlyuNie,PANi,ArMatPly1,MatPly}. 
This corresponds to supersymmetrically extended finite-gap or reflectionless systems,
which can be regarded as ``instant photos'' of solutions to the KdV equation \cite{NMPZ} and 
are characterized by the presence of a nontrivial Lax-Novikov integrals
to be operators of the odd differential order $n=2\ell+1\geq 3$ with $\ell=l+k$.
Factorization of Lax-Novikov integrals into two differential operators of orders $2l$ and $2k+1$ 
is reflected in the presence of the exotic nonlinear $\mathcal{N}=4$ Poincar\'e supersymmetry 
generated by supercharges of orders $2l$ and $2k+1$ instead of linear or nonlinear 
$\mathcal{N}=2$ Poincar\'e supersymmetry obtained usually via the Darboux, or Darboux-Crum 
 transformation construction. 
 \vskip0.1cm
 
A simple example of a system with exotic nonlinear $\mathcal{N}=4$ supersymmetry is 
 generated via the construction of Witten's supersymmetric quantum mechanics
with superpotential $W(x)= \kappa \tanh \kappa x$,
where $\kappa$ is a parameter of dimension of inverse length.
The corresponding superextended system is described by the Hamiltonian 
$\hat{\mathcal{H}}=\text{diag}\,(\hat{H}_+,\hat{H}_-)$ with $\hat{H}_-=-\frac{d^2}{dx^2}+\kappa^2$,
$\hat{H}_+=\hat{H}_--2\kappa^2/\cosh^2\kappa x$,
and first order supercharges $\hat{Q}_+=(\frac{d}{dx}-W(x))\sigma_+$, $\hat{Q}_-=(\hat{Q}_+)^\dagger$.
They generate the $\mathcal{N}=2$ Poincar\'e superalgebra
via the (anti)commutation relations
\be
\{\hat{Q}_+,\hat{Q}_-\}=\hat{\mathcal{H}}\,,\qquad
[\hat{\mathcal{H}},\hat{Q}_\pm]=0\,.
\ee
This system can also be obtained via the construction of the 
$n=2$ supersymmetry by choosing $\mathcal{W}(x)=-\frac{1}{2}\kappa\tanh\kappa x$
and $C=-\frac{1}{16}\kappa^4$ \cite{SchwAn}. 
In this case $\Delta=-\frac{\kappa^2}{\cosh^2\kappa x}(1+\frac{1}{4\sinh^2\kappa x})$, 
and the operator in the second order supercharge (\ref{QDel}) is
factorized in the form 
\be\label{2s+}
\hat{s}_+=\left(\frac{d}{dx}-\kappa\tanh \kappa x\right)\frac{d}{dx}\,.
\ee
We have here 
\be
\{\hat{S}_+,\hat{S}_-\}=\left(\hat{\mathcal{H}}-\frac{1}{2}\kappa^2\right)^2-\frac{1}{16}\kappa^4\,,\qquad
[\hat{\mathcal{H}},\hat{S}_\pm]=0\,.
\ee
The  anti-commutators 
of the first and second order supercharges generate a nontrivial 
even integral of motion, 
\be
\{\hat{S}_+,\hat{Q}_-\}=-\{\hat{S}_-,\hat{Q}_+\}=i\hat{\mathcal{L}}\,,
\ee
\be\label{L}
\hat{\mathcal{L}}=
\left(
\begin{array}{cc}
\hat{q}_+\hat{p} \,\hat{q}_+^\dagger &  0   \\
 0   &   \hat{H}_- \hat{p}
\end{array}
\right),
\ee
where $\hat{q}_+=\frac{d}{dx}-\kappa\tanh \kappa x$. Operator (\ref{L})
 satisfies the commutation relations
\be\label{center}
[\hat{\mathcal{L}},\hat{Q}_\pm]=[\hat{\mathcal{L}},\hat{S}_\pm]=[\hat{\mathcal{L}},\hat{\mathcal{H}}]=0\,,
\ee
which mean that the integral $\hat{\mathcal{L}}$ is the central element of the nonlinear superalgebra
generated by $\hat{\mathcal{H}}$, $\hat{Q}_\pm$, $\hat{S}_\pm$ and $\hat{\mathcal{L}}$.
The lower term in the diagonal 
operator $\hat{\mathcal{L}}$  is the momentum operator of a free  quantum particle multiplied by 
$\hat{H}_-$,  
while the third order differential operator $\hat{q}_+\hat{p} \,\hat{q}_+^\dagger$
is the Lax-Novikov integral of reflectionless system described by
the Hamiltonian operator $\hat{H}_+$. 
\vskip0.1cm

Operator  $\hat{\mathcal{L}}$ plays essential role
in the description of the system $\hat{\mathcal{H}}$: it detects and annihilates a unique bound state in 
the spectrum of reflectionless subsystem $\hat{H}_+$, which is  described by the wave function
$\Psi_0=\left(\sqrt{2\kappa^{-1}}\cosh\kappa x,0\right)^t$ of zero energy. 
It also annihilates the doublet of states $\Psi_+=(\tanh\kappa x,\,0)^t$ and  $\Psi_-=(0,\,1)^t$
of the system $\hat{\mathcal{H}}$ of energy $E=
\kappa^2$.  
 Besides, operator  $\hat{\mathcal{L}}$ distinguishes (with the aid of the integral $\sigma_3$)
the states
$\Psi^{\pm k}_+=(\pm ikx -\kappa\tanh\kappa x)e^{\pm ikx},\,0)^t$
and $\Psi^{\pm k}_-=(0,\, e^{\pm ikx})^t$
in the four-fold degenerate scattering part of the spectrum of $\hat{\mathcal{H}}$:
$\hat{\mathcal{L}}\Psi^{\pm k}_+=\pm k(\kappa^2+k^2)\Psi^{\pm k}_+$, 
$\hat{\mathcal{L}}\Psi^{\pm k}_-=\pm k(\kappa^2+k^2)\Psi^{\pm k}_-$.
Zero energy state $\Psi_0$ is annihilated here by all the supercharges and by 
the Lax-Novikov integral $\hat{\mathcal{L}}$,
and thus the system realizes exotic supersymmetry in the unbroken 
phase \cite{PlyuNie,ArMatPly1}.
\vskip0.1cm

Within the framework of the Darboux-Crum construction, the described 
reflectionless system $\hat{H}_+$ is obtained from the free particle system 
 $\hat{H}_0=-\frac{d^2}{dx^2}$
by taking its non-physical eigenstate $\psi_1(x)=\cosh \kappa x$ of  eigenvalue $-\kappa^2$
as the seed state  by constructing 
the operator 
\be\label{CD}
\hat{H}_+=\hat{H}_--2(\ln \W)''\,,
\ee
where $\hat{H}_-=\hat{H}_0 +\kappa^2$ and 
$\W=\psi_1(x)$. The supercharge $\hat{Q}_+$ is constructed then from the operator
$\hat{q}_+=\psi_1\frac{d}{dx}\frac{1}{\psi_1(x)}=\frac{d}{dx}-\kappa\tanh \kappa x$.
The same superpartner system $\hat{H}_+$ can be generated 
via relation (\ref{CD}) by changing  $\W=\psi_1(x)$  in it for Wronskian
of the set of eigenstates $\psi_0=1$ and $\psi_1=\sinh\kappa x$, 
which is equal, up to inessential multiplicative constant, to the same function $\W=\psi_1(x)$:
$\W(1,\sinh\kappa x)=\kappa\cosh\kappa x$.
This second DC scheme generates the intertwining operator (\ref{2s+})
corresponding to the second order supercharge $\hat{S}_+$
via the chain of relations 
$\hat{s}_+=A_2A_1$,
where $A_1=\psi_0\frac{d}{dx}\frac{1}{\psi_0}=\frac{d}{dx}$,
$A_2=(A_1\psi_1)\frac{d}{dx}\frac{1}{(A_1\psi_1)}=\hat{q}_+$.
In this construction the third order Lax-Novikov integral  $\hat{q}_+\hat{p} \,\hat{q}_+^\dagger$
of the subsystem $\hat{H}_+$ is the Darboux-dressed momentum operator of the free particle.
\vskip0.1cm

The described DC construction of superextended 
systems described by exotic $\mathcal{N}=4$ supersymmetry 
is generalized for arbitrary case of the system of the form 
$\hat{\mathcal{H}}=\text{diag}\,(\hat{H}_+,\hat{H}_-)$,
with reflectionless subsystems $\hat{H}_+$ and $\hat{H}_-$  
having  an arbitrary number and energies of bound states, but with identical  
continuous parts of their spectra \cite{ArMatPly1}.
The  key point underlying the appearance of the two supersymmerties of different orders
by means of which the partner systems $\hat{H}_+$ and $\hat{H}_-$ are related 
is that the same reflectionless system can be generated by 
two different Darboux-Crum transformations. 
One transformation is generated by the choice 
of the set of non-physical eigenstates 
\be\label{seedpsi}
\psi_1=\cosh \kappa_1 (x+\tau_1),\,
\psi_2=\sinh \kappa_2 (x+\tau_2),\,\ldots,\, \psi_n
\ee
 of the free particle system taken as the seed states.
Here $\psi_{2l+1}=\cosh \kappa_{2l+1} (x+\tau_{2l+1})$,  $\psi_{2l}=\sinh \kappa_{2l} (x+\tau_{2l})$,
$1\leq 2l< 2l+1\leq n$,   and $\kappa_j$ and $\tau_j$, $j=1,\ldots, n$, 
are arbitrary real parameters with restriction $0<\kappa_j<\kappa_{j+1}$.
  The indicated choice of eigenstates guarantees
that the Wronskian of these states takes nonzero values, and the 
potential produced via the Wronskian construction,
$V(x)=-2(\ln\W(\psi_1,\ldots,\psi_n))''$, will be nonsingular reflectionless potential
maintaining $n$ bound states. The choice of the translation parameters
$\tau_j$ in the form $\tau_j=x_{0j}-4\kappa_j^2t$  promotes the potential 
into the $n$-soliton solution to the KdV equation~\cite{AraMatPly2,Defects2}
\be
u_t=6uu_x-u_{xxx}\,.
\ee
Exactly the same reflectionless potential $V(x)$ is generated
by taking the following set of eigenstates of the free particle Hamiltonian operator,
\be\label{seedphi}
\phi_0=1,\,
\phi_1=\sinh \kappa_1 (x+\tau_1),\,
\phi_2=\cosh \kappa_2 (x+\tau_2),\,
\ldots,\, \phi_n\,,
\ee
as the seed states for the Darboux-Crum transformation.  Here
$\phi_{2l+1}=\sinh \kappa_{2l+1} (x+\tau_{2l+1})$,  
$\phi_{2l}=\cosh \kappa_{2l} (x+\tau_{2l})$,  and modulo the unimportant 
multiplicative constant, 
we have
\be\label{H01psi}
\W(\psi_1,\ldots,\psi_n)=\W(1,\psi_1'\ldots,\psi_n')\,.
\ee

\vskip0.1cm
When the number of bound states $n$ in each partner  reflectionless system 
$\hat{H}_+$ and $\hat{H}_-$ is the same but all the discrete energies 
of one subsystem are different from those of another subsystem,
one pair of supercharges will have differential order $2n$ while another pair
will have differential order $2n+1$ independently on the values of 
translation parameters
$\tau_j$ of subsystems.
This corresponds to the nature of the described Darboux-Crum transformations.
In this case one pair of the supercharges is constructed from intertwining operators 
which relate the partner system $\hat{H}_+$  via the ``virtual" free  particle system $\hat{H}_0$,
and then $\hat{H}_0$ to $\hat{H}_-$. The corresponding intertwining operators 
are composed from intertwining operators obtained from the sets of
the seed states  of the form  (\ref{seedpsi}) used for the 
construction of each partner system. Another pair of supercharges of differential order $2n+1$
is constructed from the intertwining operators of a similar form but with inserted  in the middle
free particle integral $\frac{d}{dx}$. This
corresponds to the use of the set of the seed states 
of the form (\ref{seedphi}) for one of the partner subsystems.
The Lax-Novikov integral being even generator of the exotic supersymmetry and having differential order $2n+1$
is produced  via anti-commutation of the supercharges of different differential orders.
It, however, is not a central charge of the nonlinear superalgebra:
commuting with one pair of supercharges it
transforms them into another pair of supercharges  multiplied by certain 
polynomials in Hamiltonian
$\hat{\mathcal{H}}$ of corresponding orders \cite{ArMatPly1}. 
The structure of exotic supersymmetry undergoes a reduction each time 
when some $r$ discrete energies of one subsystem coincide with any $r$ discrete energies 
of another subsystem. In this case the sum of differential orders of two  pairs of supercharges 
reduces from $4n+1$ to $4n-2r+1$, and nonlinear superlagebraic  structure acquires 
a dependence on $r$ relative translation parameters $\tau^+_j-\tau^-_{j'}$ whose indexes
$j$ and $j'$ correspond to coinciding discrete energy levels.
When all the discrete energy levels of one subsystem coincide with those
of the partner system, the Lax integral transforms into the bosonic  central charge 
of the corresponding nonlinear superalgebra \cite{ArMatPly1}. 
\vskip0.1cm

Different supersymmetric systems of the described nature 
can also be related by sending some of the translation parameters $\tau_j$ to infinity.
In such a procedure exotic supersymmetry undergoes certain transmutations, particularly, between
the unbroken and  broken phases, and 
admits  an  interpretation  in terms of 
the picture of soliton scattering \cite{AraPlyTrans}. 
\vskip0.1cm

In the interesting case of a superextended system 
unifying  two finite-gap periodic partners 
described by the associated Lam\'e potentials shifted mutually for the half of the 
period of their potentials,  the two corresponding Darboux-Crum transformations are constructed 
on the two sets of the seed states which correspond to the edges of the valence and conduction bands,
one of which is composed from periodic states while another
consists from anti-periodic states. One of such sets
corresponding to antiperiodic wave functions  has even dimension, while another,
that includes wave functions with the same period as the potentials,
has odd dimension. These sets  generate the pairs of supercharges of the correspoding even and odd
differential orders.  On these  sets of the states, certain finite-dimensional non-unitary representations 
of the $sl(2,\R)$ algebra are realized of the same even and odd dimensions
\cite{CJNP}.
Lax-Novikov integral  in such finite-gap systems with exotic nonlinear $\mathcal{N}=4$ supersymmetry 
has a nature of the bosonic central charge
and differential order equal to $2g+1$, where $g$ is the number 
of gaps in the spectrum of completely isospectral partners.
  The indicated class of the supersymmetric
finite-gap systems admits an interpretation as a planar model of a 
non-relativistic electron in periodic magnetic and electric fields that
produce a one-dimensional  crystal for two spin components 
separated by a half-period spacing \cite{CJNP}. 
Exotic supersymmetry in such systems is in  the unbroken phase 
with two ground states having the same zero energy, particularly, in the case when
one pair of the supercharges has differential order one and corresponds to the construction
of the Witten's supersymmetric quantum mechanics.  
The simplest case of such a system is given by 
the pair of the mutually shifted for the half-period
one-gap Lam\'e systems, 
\be 
\hat{H}_\pm=-\frac{d^2}{dx^2}+V_\pm(x)\,,\qquad
V_-(x)=2\text{sn}^2(x\vert k)-k^2\,,\qquad V_+(x)=V_-(x+{\bf K})\,,
\ee
where  $k$ is the modular parameter and $4{\bf K}$
is the period of the Jacobi elliptic  function $\text{sn}\,(x\vert k)$.
The extended matrix system $\hat{\mathcal{H}}$  is described by the first order supercharges 
constructed on the base of the superpotential $W(x)=-(\ln \text{dn}\,x)'$
generated by the ground state $\text{dn}\,x$ of the subsystem $\hat{H}_-$
which  has the same  period $2{\bf K}$ as the potential $V_-(x)$.
The second order supercharges are generated via the Darboux-Crum construction
on the base of  the seed states $\text{cn}\,x$ and $\text{sn}\,x$ which change sign under the shift for
$2{\bf K}$, and describe the states  
of energies $1-k^2$ and $1$ at the edges of valence and conduction bands of  $\hat{H}_-$, 
respectively. 
\vskip0.1cm

The superextended system composed from the same one-gap systems but shifted mutually
for the distance less than half-period of their potentials  is described by exotic nonlinear $\mathcal{N}=4$ 
supersymmetry with supercharges to be differential operators of the same first and second orders,
and Lax-Novikov integral having differential order three. But in this case supersymmetry is broken,
the positive energy of the doublet of the ground states depends on the value of the mutual shift,
and though the Lax-Novikov integral is the bosonic central charge,
the structure coefficients of the nonlinear superalgebra depend on the value of the shift parameter 
\cite{PANi}.
\vskip0.1cm

As was shown in \cite{BMP}, reflectionless and finite-gap periodic systems described by exotic nonlinear supersymmetry
can also be generated in  quantum  systems with a position-dependent 
mass~\cite{QuesTkach,Gang2,CCNN,coherent2}.
\vskip0.1cm

Very interesting physical properties are exhibited 
in the systems  with the exotic nonlinear $\mathcal{N}=4$ supersymmetry
realized on finite-gap systems with soliton defects \cite{Defects2,Defects1}.
By applying Darboux-Crum transformations to a Lax pair formulation of the 
KdV equation, one can construct multi-soliton solutions to this equation as well
as to the modified Korteweg-de Vries equation which 
represent different types of defects in crystalline background of the 
pulse and compression modulation types. These periodicity defects   reveal 
a   chiral asymmetry in their propagation. 
Exotic nonlinear supersymmetric structure 
in such systems  unifies solutions to the KdV and modified KdV
equations, it  detects the presence of  soliton defects in them,
distinguishes their types, and identifies the types of crystalline 
backgrounds \cite{Defects2}.

\section{Perfectly invisible $\mathcal{PT}$-symmetric 
zero-gap systems}

Darboux-Crum transformations can be realized not only on the 
base of the physical or non-physical eigenstates of  a system,
but also by including into the set of the seed states of 
Jordan and generalized Jordan states \cite{MatPly,JM2,Sch-Hal,CorJacPly,ConAst},
which, in turn, can be obtained by certain limit procedures from 
 eigenstates of a system.
 For instance one can start from the free quantum particle, and 
 choose the set of the states $(x,x^2,x^3,\ldots x^n)$,
 $x^n=\lim_{k\rightarrow 0}(\sin kx/k)^n$.
 The first state $x$ is a non-physical eigenstate of $\hat{H}_0=-\frac{d^2}{dx^2}$ 
 of zero eigenvalue. The states $x^{2l}$, $x^{2l+1}$, $l\geq 1$,  are the Jordan states 
 of order $l$ of $\hat{H}_0$: $(\hat{H}_0)^{l}$ acting on both states transforms them
 into zero energy eigenstates $\psi_0=1$ and $\psi_1=x=\widetilde{\psi_0}$, respectively.
 The Wronskian of these states is $\W(x,x^2,x^3,\ldots, x^n)=const\cdot x^n$,
 and the system generated via the corresponding Darboux-Crum transformation
 is $\hat{H}_n=-\frac{d^2}{dx^2}+\frac{n(n+1)}{x^2}$. 
 Operator $\hat{H}_n$, however, is singular on the whole real line,
 and can be identified with the Hamiltonian of the two-particle Calogero
 \cite{Calogero,Polych}
 model with the omitted center of mass coordinate, which requires 
 for definition of its domain with $x\in (0,+\infty)$ the introduction of the Dirichlet boundary condition $\psi(0^+)=0$.
 Systems $\hat{H}_0$ and $\hat{H}_n$ are intertwined
 by differential operators $\A_n=A_n\ldots A_1$ and  $\A_n^\dagger$,
 $\A_n\hat{H}_0=\hat{H}_n \A_n$,  $\A_n^\dagger\hat{H}_n=\hat{H}_0 \A_n^\dagger$
 where $A_l=\frac{d}{dx}-\frac{l}{x}$, and construction of 
 $\A_n$  corresponds to Eq. (\ref{Andef}).
The systems $\hat{H}_0$ and $\hat{H}_n$ can also be intertwined by the 
operators  $\B_n=A_n\ldots A_1A_0$ and $\B_n^\dagger$,
where $A_0=\frac{d}{dx}$, which are obtained by realizing the Darboux-Crum 
transformation constructed 
on the base of the set of the states $(x^2,\ldots,x^{n+1})$ extended with the state $\psi_0=1$.
One could take then the extended system composed from 
$\hat{H}_+=\hat{H}_n$ and $\hat{H}_-=\hat{H}_0$ with $\hat{H}_0$ restricted to the same domain 
as  $\hat{H}_n$, and construct  the supercharge operators of  differential orders $n$ and $n+1$
from  the introduced intertwining operators.
 However, we find that the supercharge constructed on the base of the intertwining operators
 $\B_n$ and $\B_n^\dagger$ will be non-physical as the intertwining 
 operator  $\B_n$ acting on physical eigenstates  $\sin kx$ of $\hat{H}_-$ of energy $k^2$
will transform them into non-physical eigenstates   $\B_n \sin kx$ of the system $\hat{H}_+$
of the same energy but not satisfying the boundary condition $\psi(0^+)=0$.
In correspondence with this, differential  operator of order $2n+1$, $\hat{\mathcal{L}}=\text{diag}\,(\hat{\mathcal{L}}_+,
\hat{\mathcal{L}}_-)$, with 
$\hat{\mathcal{L}}_+=\B_n\A_n^\dagger=\A_n \frac{d}{dx}\A_n^\dagger$ and 
$\hat{\mathcal{L}}_-= \A_n^\dagger\B_n =(\hat{H}_-)^n\frac{d}{dx}$
formally commutes with $\hat{\mathcal{H}}$,
but it is not a physical operator for the system $\hat{\mathcal{H}}$ as acting on its
physical eigenstates satisfying boundary condition at $x=0^+$, it transforms them into non-physical eigenstates
not satisfying the boundary condition.
The situation can be  ``$\mathcal{PT}$-regularized'' by shifting the variable $x$: $x\rightarrow \xi=x+i\alpha$,
where $\alpha$ is a nonzero real parameter \cite{MatPly}.
The obtained in such a way superextended system can be considered 
on the whole real line $x\in\R$, and boundary condition at $x=0$ can be omitted.
The system $\hat{H}_+(\xi)$ is 
$\mathcal{PT}$-symmetric \cite{PTSym1,PTSym2,PTSym3,DorDunTat2,PTSym4,PTSym5,PTSym6}: 
$[PT, \hat{H}_+(\xi)]=0$,
where $P$ is a  space reflection operator, $Px=-Px$, and $T$ is the operator defined by
$T(x+i\alpha)=(x-i\alpha)T$. Subsystem $\hat{H}_+(\xi)$ has one bound eigenstate of zero eigenvalue  
  described by quadratically integrable on the whole real line function $\psi_0^+=\xi^{-n}$,
 which lies at the very edge of the continuos spectrum with $E>0$.
 System $\hat{H}_+(\xi)$ therefore can be identified as $\mathcal{PT}$-symmetric zero-gap system.
 Moreover, it turns out that the transmission amplitude for this system is equal to one
 as for the free particle system, and  $\hat{H}_+(\xi)$ can be regarded as a
 perfectly invisible $\mathcal{PT}$-symmetric  zero-gap system.
 Exotic nonlinear supersymmetry of the system $\hat{\mathcal{H}}(\xi)$
 will be described by two supercharges of differential order $n$ constructed from the intertwining operators 
 $\A_n(\xi)$ and $\A_n^\#(\xi)=\A_1^\#\ldots \A_n^\#$, $\A_j^\#=-\frac{d}{dx}-\frac{j}{\xi}$,
 by supercharges of the order $n+1$ constructed from the intertwining operators
 $\B_n(\xi)$ and $\B_n^\#(\xi)$, and by the Lax-Novikov integral $\hat{\mathcal{L}}(\xi)$
 to be differential operator of order $2n+1$. 
 Operator $\hat{\mathcal{L}}(\xi)$ annihilates the unique bound state of the system 
 $\hat{\mathcal{H}}(\xi)$  and the state $\psi_0=1$ of zero energy in the spectrum of the free particle
 subsystem, and distinguishes plane waves $e^{ikx}$ in the spectrum of the free particle subsystem
 and deformed plane waves $\A_n(\xi)e^{ik\xi}$ in the spectrum of the 
 superpartner system $\hat{H}_+(\xi)$.
 \vskip0.1cm
 
 In the simplest case $n=1$, the supercharges have the form 
\be\label{Qa}
\hat{Q}_1=
\left(
\begin{array}{cc}
 0 & A_1(\xi)   \\
A^\#_1(\xi)  &  0
\end{array}
\right),\qquad
\hat{Q}_2=i\sigma_3\hat{Q}_1\,,
\ee
\be\label{Sa}
\hat{S}_1=
\left(
\begin{array}{cc}
 0 & -A_1(\xi) \frac{d}{dx}  \\
\frac{d}{dx} A^\#_1 (\xi) &  0
\end{array}
\right),\qquad
\hat{S}_2=i\sigma_3\hat{S}_1\,,
\ee
where $\hat{Q}_1=\hat{Q}_++\hat{Q}_-$,
$\hat{S}_1=\hat{S}_++\hat{S}_-$.
The Lax-Novikov integral  is
\be\label{L1def}
\hat{\mathcal{L}}=
\left(
\begin{array}{cc}
 -iA_1(\xi)\frac{d}{dx}A_1^\#(\xi) & 0  \\
0 & -i\frac{d}{dx}\hat{H}_0
\end{array}
\right).
\ee
Together with  Hamiltonian $\hat{\mathcal{H}}=\text{diag}\,(\hat{H}_1(\xi),\hat{H}_0)$
they satisfy the following nonlinear superalgebra \cite{MatPly}:
\be\label{susyHQS}
[\hat{\mathcal{H}},\hat{Q}_a]=0\,,\qquad
[\hat{\mathcal{H}},\hat{S}_a]=0\,,
\ee
\be\label{susyQQSS}
\{\hat{Q}_a,\hat{Q}_b\}=2\delta_{ab}\hat{\mathcal{H}}\,,\qquad
\{\hat{S}_a,\hat{S}_b\}=2\delta_{ab}\hat{\mathcal{H}}^2\,,
\ee
\be\label{susyQS}
\{\hat{Q}_a,\hat{S}_b\}=2\epsilon_{ab}\hat{\mathcal{L}}\,.
\ee
\be\label{susyL1QS}
[\hat{\mathcal{L}},\hat{\mathcal{H}}]=0\,,\qquad
[\hat{\mathcal{L}},\hat{Q}_a]=0\,,\qquad
[\hat{\mathcal{L}},S_a]=0\,.
\ee
\vskip0.1cm
In the case of the super-extended system 
$\hat{\mathcal{H}}=\text{diag}\,(\hat{H}_1(\xi_2),\hat{H}_1(\xi_1))$,
where $\xi_j=x+i\alpha_j$, $j=1,2$, and $\alpha_1\neq \alpha_2$,
exotic nonlinear supersymmetry is partially broken:
the doublet of zero energy bound states is annihilated by 
 the second order supercharges $\hat{S}_a$ and by the Lax-Novikov integral 
$\hat{\mathcal{L}}$, but they are  not annihilated by the first order supercharges 
$\hat{Q}_a$ \cite{MatPly}.  The first order supercharges $\hat{Q}_a$ are constructed in this case from the 
intertwining operators $A=\frac{d}{dx}+\mathcal{W}$, $\mathcal{W}=\xi_1^{-1}-\xi_2^{-1}-(\xi_1-\xi_2)^{-1}$,
and $A^\#=-\frac{d}{dx}+\mathcal{W}$.  The second order supercharges $\hat{S}_a$
are composed from
the  intertwining operators $A_1(\xi_2)A_1^\#(\xi_1)$ and $A_1(\xi_1)A_1^\#(\xi_2)$.
In the limit $\alpha_1\rightarrow \infty$, the system 
$\hat{\mathcal{H}}=\text{diag}\,(\hat{H}_1(\xi_2),\hat{H}_1(\xi_1)$
transforms into the system given by the $\mathcal{PT}$-symmetric 
Hamiltonian $\hat{\mathcal{H}}=\text{diag}\,(\hat{H}_1(\xi_2),\hat{H}_0)$,
and exotic nonlinear supersymmetry in the partially broken phase transmutes into the supersymmetric
structure corresponding to the unbroken phase \cite{MatPly}.
\vskip0.1cm

It is interesting to note that if to use the appropriate linear combinations of the 
Jordan states of the quantum free particle as the seed states  for the Darboux-Crum transformations,
one can construct $\mathcal{PT}$-symmetric  time-dependent potentials which will satisfy equations of the KdV hierarchy 
and will exhibit  a behaviour typical for extreme (rogue) waves \cite{MatPly}.

\section{Nonlinear superconformal symmetry of  the $\mathcal{PT}$-symmetric  zero-gap  Calogero  systems}
Free particle system is characterized by the Schr\"odinger symmetry 
generated
by the first order integrals $\hat{P}_0=\hat{p}=-i\frac{d}{dx}$ and $\hat{G}_0=x+2it\frac{d}{dx}$,
and the second order integrals  $\hat{H}_0=-\frac{d^2}{dx^2}$,
$\hat{D}_0=\frac{1}{4}\{\hat{P}_0,\hat{G}_0\}$ and 
$\hat{K}_0=\hat{G}_0^2$. Operators  $\hat{G}_0$ as well as 
$\hat{D}_0$ and $\hat{K}_0$ are dynamical integrals of motion satisfying 
the equation of motion of the form $\frac{d}{dt}\hat{I}=\frac{\partial}{\partial t}\hat{I}-[\hat{H}_0,\hat{I}]=0$.
These time-independent and dynamical  integrals generate the 
Schr\"odinger algebra
\begin{eqnarray}\label{sl(2,R)}
&[\hat{D}_0,H_0]=i\hat{H}_0\,,\qquad
[\hat{{D}}_0,\hat{K}_0]=-i\hat{K}_0\,,\qquad
 [\hat{K}_0,\hat{H}_0]=8i\hat{{D}}_0\,,&\\
\label{D_0F}
&[\hat{{D}}_0,\hat{P}_0]=\frac{i}{2}\hat{P}_0\,,\qquad 
\hat[{D}_0,\hat{G}_0]=-\frac{i}{2}\hat{G}_0\,,&\\
\label{H_0F}
&[\hat{H}_0,\hat{G}_0]=-2i\hat{P}_0\,,
\qquad
[\hat{H}_0,\hat{P}_0]=0\,,&\\
\label{K_0F}
&[\hat{K}_0,\hat{P}_0]=2i\hat{G}_0\,,\qquad
[\hat{K}_0,\hat{G}_0]=0\,,&\\
\label{G_0P_0}
&[\hat{G}_0,\hat{P}_0]=i\,\mathbb{I} \,.&
\end{eqnarray}
Eqs. (\ref{sl(2,R)}) and  (\ref{G_0P_0})  correspond to the $sl(2,\R)$ 
and Heisenberg subalgebras, respectively. 
If we make a shift $x\rightarrow \xi=x+i\alpha$,
and make Darboux-dressing of operators 
$\hat{P}_0$, $\hat{G}_0$,
$\hat{D}_0$ and $\hat{K}_0$, we find the integrals of motion 
for the perfectly invisible zero-gap  $\mathcal{PT}$-symmetric system $\hat{H}_1(\xi)$.
These are $\hat{P}_1(\xi)=A_1(\xi)\hat{P}_0A_1^\#(\xi)$,
$\hat{G}_1(\xi)=A_1(\xi)\hat{G}_0A_1^\#(\xi)$, and 
\be
\hat{D}_1(\xi)=-\frac{i}{2}\left(\xi\frac{d}{dx}+\frac{1}{2}\right)
-t\hat{H}_1(\xi)\,,
\ee
\be
\hat{K}_1(\xi)=\xi^2-8t\hat{D}_1(\xi)-4t^2 \hat{H}_1(\xi)\,,
\ee
where the dynamical integrals 
$\hat{D}_1(\xi)$ and $\hat{K}_1(\xi)$
have been extracted from the corresponding Darboux-dressed operators
by omitting in them the operator factor $\hat{H}_1(\xi)$ \cite{JM2}.
Operators  $\hat{H}_1(\xi)$,  $\hat{D}_1(\xi)$ and $\hat{K}_1(\xi)$
generate the same $sl(2,\R)$ algebra as in the case of the free particle.
But now we have relations
\be
\label{D_1F}
[{D}_1,{P}_1]=\frac{3}{2}i{P}_1\,,\qquad
[{D}_1,G_1]=\frac{i}{2}G_1\,,\qquad
[K_1,{P}_1]=6iG_1\,,
\ee
\be\label{G_1P_1}
[G_1,{P}_1]=3i(H_1)^2\,
\ee
instead of the corresponding relations of the free particle system.
In addition, two new dynamical integrals of motion,
\be
V_1(\xi)=i\xi^2A^\#_1(\xi)-4tG_1(\xi)-4t^2{P}_1(\xi)\,
\ee
and
\be
R_1(\xi)=\xi^3-6tV_1(\xi)-12t^2G_1(\xi)-8t^3{\xi}_1\,,
\ee
are  generated via the commutation relations 
\be\label{Lambda}
[\hat{K}_1,\hat{G}_1]=-4i\hat{V}_1\,,\qquad
\qquad
[\hat{K}_1,\hat{V}_1]=-2i\hat{R}_1\,,
\ee
and we obtain additionally the commutation relations  
\begin{eqnarray}
&[\hat{V}_1,\hat{H}_1]=4i\hat{G}_1\,,\qquad
[\hat{V}_1,\hat{D}_1]=\frac{i}{2}\hat{V}_1\,,&\\
\label{V_1quad}
&[\hat{V}_1,\hat{P}_1]=12i\hat{H}_1\hat{D}_1-6\hat{H}_1\,,\qquad
[\hat{V}_1,\hat{G}_1]=12i (\hat{D}_1)^2 +\frac{3}{4}i\, \mathbb{I} \,,&\\
&[\hat{R}_1,\hat{H}_1]=6i\hat{V}_1\,,\qquad
[\hat{R}_1,\hat{D}_1]=\frac{3}{2}i\hat{R}_1\,,
\quad
[\hat{R}_1,\hat{K}_1]=0\,,&\\
\label{R_1quad}
&[\hat{R}_1,\hat{P}_1]=36i\,\hat{D}_1^2+\frac{21}{4}i\, \mathbb{I} \,,\quad
[\hat{R}_1,\hat{G}_1]=12i\,\hat{D}_1\hat{K}_1-6\hat{K}_1\,,\quad
[\hat{R}_1,\hat{V}_1]=3i\,\hat{K}_1^2\,.&
\end{eqnarray}
The Schr\"odinger algebra of the free particle 
is extended for its nonlinear generalization in the case of the 
$\mathcal{PT}$-symmetric system 
$\hat{H}_1(\xi)$, which is generated by the operators 
$\hat{H}_1(\xi)$, $\hat{P}_1(\xi)$, $\hat{G}_1(\xi)$,
$\hat{D}_1(\xi)$, $\hat{K}_1(\xi)$, $\hat{V}_1(\xi)$, $\hat{R}_1(\xi)$
and central charge $\mathbb{I}$ (equals to $1$ in the chosen system of units).
All these integrals are eigenstates of the dilatation operator 
$\hat{D}_1(\xi)$
with respect
to its adjoint action. 
\vskip0.1cm
Now we can consider the generalized and extended superconformal symmetry
of the system described by the matrix Hamiltonian operator 
$\hat{\mathcal{H}}=\text{diag}\,(\hat{H}_1(\xi),\hat{H}_0)$.
Supplying the Hamiltonian $\hat{\mathcal{H}}$
and Lax-Novikov integral (\ref{L1def}) 
with the bosonic integrals
$\hat{\mathcal{D}}=\text{diag}\,(\hat{D}_1(\xi),\hat{D}_0(\xi))$,
$\hat{\mathcal{K}}=\text{diag}\,(\hat{K}_1(\xi),\hat{K}_0(\xi))$,
and commuting them with supercharges 
(\ref{Qa}) and (\ref{Sa}),
we obtain a  nonlinear 
superalgebra that describes the symmetry of the system $\hat{\mathcal{H}}$, which 
corresponds to some
nonlinear extension of the super-Schr\"odinger algebra.
It is generated by the set of the even (bosonic)  integrals 
$\hat{\mathcal{H}}$, 
$\hat{\mathcal{D}}$, $\hat{\mathcal{K}}$,
$\hat{\mathcal{L}}$, $\hat{\mathcal{G}}$, 
$\hat{\mathcal{V}}$, $\hat{\mathcal{R}}$, $\hat{\mathcal{P}}_-$,
$\hat{\mathcal{G}}_-$, $\Sigma=\sigma_3$,
$\hat{\mathcal{I}}=\text{diag}\,(1,1)$,
and by  the    odd (fermionic)   integrals 
$\hat{\mathcal{Q}}_a$, $\hat{\mathcal{S}}_a$,  and 
 $\hat{\lambda}_a$, $\hat{\mu}_a$ and $\hat{\kappa}_a$, $a=1,2$,
where
\begin{eqnarray}\label{G(x)}
&\hat{\mathcal{G}}=\text{diag}\,\left(\hat{G}_1(\xi),\,  \frac{1}{2}\{\hat{G}_0(\xi),\hat{H}_0\}\right)\,,\qquad
\mathcal{V}=
i\xi^2A_1^{\alpha\#}\mathcal{I}-4t\mathcal{G}-4t^2\mathcal{L}\,,&
\end{eqnarray}
 \be\label{mathcalR}
\hat{\mathcal{R}}=
\xi^3\mathcal{I} -6t\hat{\mathcal{V}} -12t^2\hat{\mathcal{G}} -8t^3\hat{\mathcal{L}}\,,
\ee
\begin{eqnarray}\label{mathP-}
&\hat{\mathcal{P}}_-=\frac{1}{2}(1-\sigma_3)\hat{P}_0\,,\qquad
\hat{\mathcal{G}}_-=\frac{1}{2}(1-\sigma_3)\hat{G}_0(\xi),&
\end{eqnarray}
 \be\label{lambda}
\hat{\lambda}_1=
\left(
\begin{array}{cc}
 0& i\xi  \\
-i\xi &  0
\end{array}
\right)
 -2t\hat{\mathcal{Q}}_1\,,\qquad 
 \hat{\lambda}_2=i\sigma_3\hat{\lambda}_1\,,
\ee
 \be\label{mu1}
\hat{\mu}_1=
\left(
\begin{array}{cc}
 0& \xi \hat{P}_0 \\
\hat{P}_0 \xi  &  0
\end{array}
\right) -2t\hat{\mathcal{S}}_1\,,\qquad \hat{\mu}_2=i\sigma_3\hat{\mu}_1\,,
\ee
\be\label{kappa}
\hat{\kappa}_1=
\left(
\begin{array}{cc}
 0& \xi^2  \\
\xi^2  &  0
\end{array}
\right)
 -4t\hat{\mu}_1-4t^2\hat{\mathcal{S}}_1\,,\qquad 
 \hat{\kappa}_2=i\sigma_3\hat{\kappa}_1\,,
\ee
and we use the notation $\hat{G}_0(\xi)=\hat{G}_0(x+i\alpha)$.
Explicit form of the nonlinear superalgebra generated by these integrals 
of motion of the system $\hat{\mathcal{H}}$
is presented in \cite{JM2}.
All the even and odd integrals here are eigenstates
of the matrix dilatation operator $\hat{\mathcal{D}}$.
\vskip0.1cm

Essentially different generalized nonlinear superconformal structure
appears in the system described by the matrix Hamiltonian
$\hat{\mathcal{H}}=\text{diag}\,(\hat{H}_1(\xi_2),\hat{H}_1(\xi_1))$
and characterized by the partially broken exotic nonlinear $\mathcal{N}=4$ supersymmetry.
In that case the number of the even and odd integrals of motion  is the same 
as in the system $\hat{\mathcal{H}}=\text{diag}\,(\hat{H}_1(\xi),\hat{H}_0)$
in the phase with unbroken supersymmetry.
However, no odd (fermionic) integral of motion is eigenstate 
of the matrix dilatation operator $\hat{\mathcal{D}}=\text{diag}\,(\hat{D}_1(\xi_2),\hat{D}_1(\xi_1))$,
and, as a result, the structure of the nonlinear superalgebra has more complicated form.
When one the shift parameters,  $\alpha_1$,   is sent to infinity,
the system $\hat{\mathcal{H}}=\text{diag}\,(\hat{H}_1(\xi_2),\hat{H}_1(\xi_1))$ 
transforms into the system  $\hat{\mathcal{H}}=\text{diag}\,(\hat{H}_1(\xi),\hat{H}_0)$
in the unbroken phase of the exotic nonlinear $\mathcal{N}=4$ super-Poincar\'e symmetry, 
and all the integrals of the latter system can be reproduced from the integrals 
 of the former system. The relation between
the integrals turns out to be  rather non-trivial
and requires some sort of  a ``renormalization" \cite{JM2}.

\section{Rationally extended harmonic oscillator and conformal mechanics systems}
Quantum harmonic oscillator (QHO) and conformal mechanics
systems \cite{deAFF}--\cite{BonCorLatWal}
described by de Alfaro-Fubini-Furlan (AFF) model 
\cite{deAFF}
are characterized by conformal symmetry. 
In the  case of harmonic oscillator, 
like in the free particle case,  
 it extends to the Schr\"odinger symmetry \cite{Nied,BecHusSQHO,BDH,BDH2}.
Heisenberg subalgebra in the free particle system
is generated by the momentum operator being time-independent 
integral of motion, and by generator 
of the Galilean boosts $\hat{G}_0$, which is a dynamical integral of  motion.
In the case of the QHO, Heisenberg subalgebra is generated by two dynamical integrals 
of motion to be linear in  the ladder operators. In correspondence with 
this, ladder operators are the spectrum-generating operators of the QHO
having discrete equidistant spectrum instead of the continuos spectrum of the 
free particle. As a consequence of these similarities and differences
between the free particle and QHO,
exotic supersymmetry can  also be generated by 
Darboux-Crum transformations applied
to the latter system. 
Instead of the two pairs of   time-independent supercharge generators 
in superextended reflectionless systems, 
in superextended systems constructed from the pairs
of the rational extensions of the QHO, only two supercharges
are time-independent integrals, while
other two  odd generators are dynamical integrals of motion.
As a result, instead of the exotic nonlinear $\mathcal{N}=4$ supersymmetry
of the paired reflectionless (and finite-gap) systems,
 in the case of the deformed oscillator systems there appear
 some nonlinearly deformed and generalized  super-Schr\"odinger 
 symmetry. The super-extended systems composed from 
 the AFF model (with special values $g=n(n+1)$ of the coupling constant 
in its  additional potential term  $g/x^2$) and  its rational extensions
are described by the nonlinearly deformed and generalized 
superconformal symmetry \cite{InzuPlyHidconf}. 
\vskip0.1cm

 Let us consider first in more detail the case of rational deformations of the QHO 
 system \cite{Dubov,CarPly1,InzuPlyHidconf,CIPConf,CPRS,FelSmi,GUKM1,MarQue1,MarI}.
 To generate a rational deformation of the QHO, it is necessary to choose the set of
 its physical or non-physical eigenstates as seed states for the Darboux-Crum 
 transformation so that their Wronskian will take nonzero values.
 In this way we generate an almost isospectral quantum system 
 with difference only in finite number of added or eliminated energy levels.
 The QHO Hamiltonian $\hat{H}_{osc}=-\frac{d^2}{dx^2}+x^2$
 possesses the same symmetry under the Wick rotation as the quantum free particle system:
 if $\psi(x)$ is a solution of the time-independent Schr\"odinger equation
 $\hat{H}_{osc}(x)\psi(x)=E\psi(x)$,
 then $\psi(ix)$ is a solution of equation 
  $\hat{H}_{osc}(x)\psi(ix)=-E\psi(ix)$.
  To construct a rational deformation of 
  the QHO described by a nonsingular on the whole real line potential,
  one can take the following set 
  of the non-physical eigenstates of   $\hat{H}_{osc}$
  as the seed states for the Darboux-Crum transformation:
  \be\label{set-}
  (\psi^-_{j_1},\ldots,\psi^-_{j_1+l_1}), \,\,
   (\psi^-_{j_2},\ldots,\psi^-_{j_2+l_2}),\,\, \ldots,\,\,
   (\psi^-_{j_r},\ldots,\psi^-_{j_r+l_r}),
   \ee
   where $j_1=2g_1$, 
   $j_{k+1}=j_{k}+l_k+2g_{k+1}$, $g_k=1,\ldots,$
   $l_k=0,1,\ldots$, $k=1,\ldots,r-1$.
   Here  $\psi^-_n(x)=\psi_n(ix)$, $n=0,\ldots$,  is a non-physical eigenstate
   of $\hat{H}_{osc}$ of eigenvalue $E^-_n=-(2n+1)$,
   obtained by Wick rotation from a (non-normalized) physical eigenstate
   $\psi_n(x)=H_n(x)e^{-x^2/2}$ of energy $E_n=2n+1$, 
   where $H_n(x)$ is Hermite polynomial of order $n$.
   The indicated set of non-physical eigenstates of 
  $\hat{H}_{osc}$ guarantees that the Wronskian
 of the chosen seed states,
 $\W=\W(-n_m,\ldots,-n_1)$, takes nonzero values for all $x\in\R$ 
 \cite{CarPly}.
 Here we assume that $n_m>\ldots>n_1>0$,
 and in what follows we use 
  the notation for physical and non-physical eigenstates 
$n=\psi_n$ and  $-n=\psi^-_n$, respectively.
 The DC scheme based on the set of the non-physical states 
having  negative eigenvalues was  called  ``negative" in \cite{CIPConf}.
Wronskian $\W=\W(-n_m,\ldots,-n_1)$ is equal to 
some polynomial multiplied by $\exp(n_-x^2/2)$, 
where $n_-=(l_1+1)+\ldots +(l_r+1)$ is the number of the chosen seed states,
and according to Eq. (\ref{V-V+DC}), the DC transformation
generates the system described by the harmonic term $x^2$ 
extended by some  rational in $x$ term.  
Transformation based on the negative scheme $(-n_m,\ldots,-n_1)$ 
introduces effectively into the spectrum of the QHO the  $n_-$ bound states
of energy levels $-2n_m-1$, $\ldots$, $-2n_1-1$.
These additional energy levels are grouped into $r$ ``valence" bands with $l_k+1$ 
levels in the band with index $k$, which are separated by gaps of the size $4g_k$,
with the first valence band separated from the infinite equidistant part of the spectrum
by the gap of the size $4g_1$.  
The same structure of the spectrum can be achieved alternatively by
eliminating $n_+=2(g_1+\ldots+g_r)$ energy levels from the spectrum of the QHO
by  taking $n_+$ physical states 
\be\label{set+}
(\psi_{l_r+1},\ldots,\psi_{l_r+2g_r}),\,\,\ldots,\,\,
 (\psi_{n_m-2g_1+1},\ldots,\psi_{n_m}),
 \ee
 organized into $n_-$ groups.
 \vskip0.1cm
 
 The duality of the negative and positive schemes based on the 
 sets of the seed states (\ref{set-}) and (\ref{set+}) can be established as follows.
Applying Eq.  (\ref{Wtiltil}) with $\psi_*=-0$, 
and equalities  $\psi^-_0\frac{d}{dx}\frac{1}{\psi^-_0}=-a^+$,
$a^+\widetilde{\psi^-_0}=\psi_0$,
$a^+(-n)=-(n-1)$,
where $a^+=-\frac{d}{dx}+x$ is the raising  ladder operator
of the QHO,
we obtain the relation \cite{CIPConf}
\be\label{Hosc-0}
\W(-n_m,\ldots,-n_1)=\W(-0,\widetilde{-0},-n_m,\ldots,-n_1)=e^{x^2/2}\W(0,-(n_m-1),-(n-1)).
\ee
It means that the negative scheme
generated by the set of the $n_-$ non-physical seed states
$(-n_m,\ldots,-n_1)$ and the ``mixed" scheme 
based on the set of the seed states $(0,-(n_m-1),-(n-1))$
involving  the ground eigenstate 
generate, according to Eq.
(\ref{V-V+DC}),  the same quantum system 
but given by the Hamiltonian operator shifted for the additive constant term:
the potential obtained on the base of the indicated 
mixed scheme will be shifted  for the constant $+4$
in comparison with the potential generated via the DC transformation
based on the negative scheme. 
Eq. (\ref{Hosc-0}) is analogous to the Wronskian relation (\ref{H01psi})
for the  free particle states,  with the state $\psi_0=1$ and operator
 $\psi_0\frac{d}{dx}\frac{1}{\psi_0}=\frac{d}{dx}$
there to be analogous to the ground state and raising  ladder operator
of the QHO here.
In (\ref{H01psi}), however, the Wronskian equality does not 
contain any nontrivial functional factor in comparison with the 
exponential multiplier appearing in (\ref{Hosc-0}).
As a result, as we saw before,    in the case of the free particle 
any reflectionless system can be generated from it
by means of the two DC transformations, which produce exactly the same 
Hamiltonian operator. Consequently, we construct  there
two pairs of the supercharges  for the corresponding
super-extended system which are the integrals of motion 
not depending explicitly on time.  
On the other hand, in the case of 
a super-extended system produced  from the QHO we shall have
two fermionic integrals to be true, time-independent integrals of motion,
but  two other odd generators of the superalgebra will be time-dependent,
dynamical integrals of motion.
\vskip0.1cm

Applying repeatedly the procedure of  Eq. (\ref{Hosc-0}),
we obtain finally the relation~\cite{CIPConf}
\be\label{W-W+}
\W(-n_m,\ldots,-n_1)=e^{(n_m+1)x^2/2}\W(n_1',\ldots, n'_m=n_m),
\ee
where $0<n'_1<\ldots<n'_m=n_m$.
This relation means that the negative scheme $(-n_m,\ldots,-n_1)$
with $n_-$  seed states is dual to the positive  scheme
$(n_1',\ldots, n'_m=n_m)$ with $n_+=n_m+1-n_-=2(g_1+\ldots+g_k)$ seed states
representing physical eigenstates of the QHO.
The two dual schemes can be unified in one ``mirror" diagram, in which
any of the two schemes can be obtained from another by a kind of a 
``charge conjugation", see ref. \cite{CIPConf}. In this way we obtain, as an example, 
the pairs of dual schemes $(-2)\sim (1,2)$ and
$(-2,-3)\sim (2,3)$.
Eq. (\ref{W-W+}) means that the dual schemes generate the same 
rationally extended QHO system but the Hamiltonian
corresponding to the positive scheme will be shifted in comparison to the Hamiltonian
produced on the base of the negative scheme for additive constant equal to
$2(n_++n_-)=2(n_m+1)$.
One can also note that in comparison with the 
free particle case, the total number of the seed states in both 
dual schemes 
can be odd or even.
\vskip0.1cm

We denote by $\A_{(-)}^-$
the intertwining operator $\A_{n_-}$ constructed on the base 
of the negative scheme, 
and $\A_{(-)}^+\equiv (\A_{(-)}^-)^\dagger$,
see Eq. (\ref{Andef}). These are differential operators 
of order $n_-$. Analogously, the intertwining operators constructed 
by employing the dual positive scheme we denote as
$\A_{(+)}^-$,
and $\A_{(+)}^+\equiv (\A_{(+)}^-)^\dagger$; they
are differential operators of order $n_+$.
We denote by $\hat{L}_{(-)}$ and $\hat{L}_{(+)}$ the Hamiltonian operators generated  
from the QHO Hamiltonian $\hat{H}_-=\hat{H}_{osc}$
by  means of the DC transformation realized  
on the base of  the negative and positive dual schemes, 
respectively. Then $\hat{L}_{(+)}=\hat{L}_{(-)}+2(n_++n_-)$, 
$\A_{(-)}^-\hat{H}_-=\hat{L}_{(-)}$, $\A_{(+)}^-\hat{H}_-=\hat{L}_{(+)}$.
For the rationally deformed QHO system $\hat{L}_{(-)}$
one can construct three pairs of the ladder operators, two of which 
are obtained by Darboux-dressing of the ladder operators 
of the QHO system,
$\mathcal{A}^\pm=\A_{(-)}^-a^\pm \A_{(-)}^+$,
and $\mathcal{B}^\pm=\A_{(+)}^-a^\pm \A_{(+)}^+$,
while the third pair is obtained by gluing 
different intertwining operators, 
$\mathcal{C}^-=\A_{(+)}^- \A_{(-)}^+$,
$\mathcal{C}^+=\A_{(-)}^- \A_{(+)}^+$.
These  ladder operators detect all the separated states in the rationally deformed QHO system
$\hat{L}_{(-)}$ (or $\hat{L}_{(+)}$)
organized into the valence bands; they also  distinguish the valence bands themselves,
and any of the the two sets ($\mathcal{C}^\pm, \mathcal{A}^\pm)$ or 
($\mathcal{C}^\pm, \mathcal{B}^\pm)$ represents  the complete spectrum-generating 
set of the ladder operators of the system $\hat{L}_{(-)}$. 
The explicitly depending on time operators  
$\mathcal{A}^\pm e^{\mp 2it}$, $\mathcal{B}^\pm e^{\mp 2it}$
and $\mathcal{C}^\pm e^{\pm2(n_++n_-) it}$ are the dynamical integrals of motion
of the system $L_{(-)}$ which, being higher derivative differential operators,
have a nature of generators of a hidden symmetry.
If we construct now the extended system $\hat{\mathcal{H}}=\text{diag}\,(\hat{L}_{(-)}, \hat{H}_-)$,
the pair of the supercharges  constructed from the intertwining operators
$\A_{(-)}^\pm$ will be its time-independent odd integrals of motion,
while from the intertwining operators $\A_{(+)}^\pm$
we obtain a pair of the fermionic dynamical integrals of motion.
Proceeding from these  odd integrals of motion and the 
Hamiltonian $\hat{\mathcal{H}}$, one can generate a nonlinearly
deformed generalized super-Schr\"odinger symmetry of the 
super-extended system $\hat{\mathcal{H}}$.
In  the super-extended system $\hat{\mathcal{H}}=\text{diag}\,(\hat{L}_{(+)}, \hat{H}_-)$,
the pair of the time-independent supercharges is constructed from the
pair of intertwining operators $\A_{(+)}^\pm$,
while the dynamical fermionic integrals of motion are obtained 
from the intertwining operators $\A_{(-)}^\pm$.
This picture with the nonlinearly
deformed generalized super-Schr\"odinger symmetry
can also be extended for the case of a super-extended 
system $\hat{\mathcal{H}}$ composed from any pair of the rationally deformed 
quantum harmonic oscillator systems.

In \cite{CIPConf}, it was shown that the AFF model
 $\hat{H}_g=-\frac{d^2}{dx^2}+x^2+\frac{g}{x^2}$
with 
special values $g=n(n+1)$ of the coupling constant
can be obtained by applying the appropriate CD transformation
to the half-harmonic oscillator obtained from the QHO by introducing 
the infinite potential barrier at $x=0$. 
As a consequence, rational  deformations of the  AFF
conformal mechanics model can be obtained 
by employing some modification of the described 
DC transformations based on the dual schemes applied to the QHO system.
The corresponding super-extended systems composed from rationally 
deformed versions of the conformal mechanics are described by 
the nonlinearly
deformed generalized superconformal symmetry
instead of the nonlinearly
deformed generalized super-Schr\"odinger symmetry
appearing in the case of the super-extended rationally
deformed QHO systems, see \cite{InzuPlyHidconf}.
The construction of rational deformations for the
AFF model can be generalized for the case of arbitrary 
values of the coupling constant $g=\nu(\nu+1)$ \cite{InzPlyNu}.

\section{Conclusion}

We considered nonlinear supersymmetry of one-dimensional mechanical systems which has the nature  of  
the hidden symmetry  generated by  supercharges of higher order  in momentum. 
In the case of reflectionless, finite-gap, rationally deformed 
oscillator and conformal mechanics systems, as well as in a special class of 
the $\mathcal{PT}$-regularized
Calogero systems, the nonlinear $\mathcal{N}=2$ Poincar\'e  supersymmetry
 expands up to   exotic nonlinear $\mathcal{N}=4$ supersymmetric 
and nonlinearly deformed generalized super-Schr\"odinger 
or superconformal structures.

Classical symmetries described 
by the linear Lie algebraic structures are promoted by 
geometric quantization  to the quantum level \cite{GeoQua1,GeoQua2}. 
Though nonlinear symmetries described  $W$-type algebras
can be produced from linear symmetries via some reduction 
procedure \cite{BoerHarTji}, 
the problem of generation of non-linear quantum 
mechanical supersymmetries from the linear ones was not studied 
in a systematic way.
 It would be interesting to investigate this problem
bearing particularly   in mind the problem 
of the quantum anomaly associated with nonlinear 
supersymmetry \cite{KliPly}.  
Some first steps were realized in this direction in \cite{SchwAn}
in the light of the so called coupling constant metamorphosis
mechanism \cite{Hieat}. 
Note also that, as was shown in \cite{PlyPara},  nonlinear
supersymmetry of purely parabosonic systems
can be obtained by reduction of parasupersymmetric systems.
\vskip0.1cm
 
Hidden symmetries can be associated with 
the presence of the peculiar geometric structures 
in the corresponding  systems  \cite{Cariglia,Frolov,CGGH}.
It would be interesting to investigate 
nonlinear supersymmetry and related 
exotic nonlinear supersymmetric and 
superconformal structures from a similar perspective.

\vskip0.3cm

\noindent {\bf Acknowledgements}
\vskip 0.2cm
\noindent
Financial support from  research projects  Convenio Marco Universidades del Estado (Project USA1555), Chile,  
and MINECO (Project MTM2014-57129-C2-1-P), Spain, is acknowledged.

%%%%%%%%%%%%%%%%%%%%%%%%%%%%%%%%%%%%%%%%%%%%%%%%%%%%%%%%%%%%
%%%%%%%%%%%%%%%%%%%%%%%%%%%%%%%%%%%%%%%%%%%%%%%%%%%%%%%%%%%%

\end{document}